\documentclass[10pt]{article}
\usepackage{amsfonts,amssymb}
\def\BB{\mathfrak B}
\def\diag{\mathop{\mathrm{diag}}\nolimits}

\begin{document}
\title {Small Perturbations in General Relativity:\\
        Tensor Harmonics of Arbitrary Symmetry}
\author{R.A. Konoplya \\
Department of Physics, Dnepropetrovsk State University, \\
per. Nauchny 13, Dnepropetrovsk, 49050 Ukraine}
\date{\today}
\maketitle
\begin{abstract}
We develop a method for constructing of the basic functions
with witch to expand small perturbations of space-time in General
Relativity. The method allows to obtain the  tensor harmonics for
perturbations of
the background space-time admitting an arbitrary group of isometry,
and to split the linearized Einstein equations
into the irreducible combinations.
The essential point of the work is the construction of the
generalized Casimir operator for the underlying group,
which is defined not only on vector but
also on tensor fields. It is used to construct  the basic functions for
spaces of tensor representations of the background metric's group of
isometry.
The method, being general, is applied here to construction
of the basic functions for the case of the three-parameter
group of isometry
$G_3$ acting on the two-dimensional non-isotropic surface of transitivity.
As quick illustrations of the method we consider the well-known
particular
cases: cylindrical harmonic for  the flat  space-time, and Regge-Wheller
spherical harmonics for the Schwarzschild metric.
\end{abstract}

\section{Introduction}

There are the two widely accepted ways to construct tensor
harmonics. The first is based on
the construction of the tensor
basis, corresponding to a given underlying symmetry of space
\cite{Wigner}. The second way is to consider the action of the
complete set of the invariant commutative operators on  scalar
harmonics of space. However each of these methods, being well
developed in General Relativity
for spaces of high symmetries, such as spherical \cite{Reg},
\cite{Lif} for example,
encounters serious difficulties when dealing with spaces of low
symmetry. In the last few years the perturbation formalism in General
Relativity has attracted some attention for a possibility of
detecting gravitational waves from astrophysical sources by
antennas. This stimulated the development of the second order
perturbation formalism for Schwarzschild black holes and, in this connection,
a review of the first one (see \cite{Price1} and references therein).

In the present work we propose the method of the
obtaining tensor harmonics that enormously simplifies calculations
for first order perturbations in General Relativity and promises some
advantages when
dealing with the second order. In  \cite{Gl1},
it was proposed the natural
generalization of the Casimir operator defined so that
the Lie derivatives are put instead of the corresponding infinitesimal
operators (see formula (\ref{7}) of the present work).
The metric used in the definition admits a group of isometry stipulated
by the invariance condition for this operator.
The generalized Casimir operator is invariant in the space of tensor
functions. This made it possible to formulate correctly and solve the
eigenvalue and tensor eigenfunction problems, and, thereby, to construct
tensor representations of various groups. However, if we follow this
approach "directly", the system of differential equations generated by a
given generalized Casimir operator
appears with "tangled" components of tensor functions,
and thus the direct solving of the tensor eigenfunction problem, in
general case becomes impossible. An accurate and natural way to
"disentangle" such system of
differential equations (i.e. to separate variables)
is to use invariant technique developed when considering
decomposition of the tangent bundle of pseudo-Riemannian manifolds
into the corresponding subbundles \cite{Gl-Kon2}.

When at last the set of basic functions for a given symmetry of the
background metric are obtained we can expand the small perturbations of
a metric in terms of these basic functions, substitute such an expanded
metric for that into the linearized Einstein equations, and try to separate
variables into them. For an Abelian
underlying group of symmetry owing to the Lie theorem it becomes
obvious that the splitting of the linearized Einstein equations
into irreducible combinations eventually leads to separations of
the corresponding variables. However in the general case we can
not know beforehand whether variables will  be separated or not.

The paper is organized as follows. In Sec.II we give, briefly, the method
for constructing of tensor harmonics and for obtaining of the
linearized Einstein equations in form of irreducible combinations.
In Sec.III the tensor harmonics are obtained for the case of
three-parameter group on the two-dimensional non-isotropic surface of
transitivity. Sec.IV deals with the two particular cases of the previous
section: cylindrical waves as perturbations of the flat space-time
and spherical Regge-Wheeler waves.

\section{The method}

Let $M^n$ be  an $n$-dimensional manifold and $G^r$ is an $r$-parameter
transformation group on $M^n$. Note, that unlike the Lie algebra of the
tangent to one-parameter subgroups of the group  $G^r$
vectors $\xi_i$, defined only on functions, that of the
Lie derivatives
defined on tensors of an arbitrary type, are invariant
operators $L_{\xi_i}$ under the general coordinate
transformations.
Hence it seems
natural to define the generalized Casimir operator of the second order as
\begin{equation}\label{7}
        G=\gamma^{ik}L_{\xi_i}L_{\xi_k},
\end{equation}
where $\gamma^{ik} \: (i,k=1,2,...r)$ are contravariant components of some unknown
ù metric, which is subject to be determined. Since this operator
commutes with all the operators $L_{\xi_i}$ of the
representation we have
\begin{equation}\label{8}
(L_{\xi_j}\gamma^{-1})^{ik}\equiv \xi_j \gamma^{ik}+C^i_{jl}\gamma^{lk}+C^k_{jl}\gamma^{il}=0,
\end{equation}
where the tensor
\begin{equation}\label{9}
\gamma^{-1}=\gamma^{ik} \xi_i \otimes\xi_k \qquad (rank\,||g^{-1}||=s)
\end{equation}
defines the metric on covectors belonging to the surfaces of transitivity
$M^s\subset M^n  (s\! \leq\! r,  ~s\! \leq\! n)$. Owing to (\ref{8}),
the group $G^r$ on the surfaces of transitivity is
a group of isometries, where $\vec{\xi_i}$ are the Killing's vectors.
Solutions of the Killing's equations (\ref{8})
give us the metric for the definition of the generalized Casimir operator
(\ref{7}).
It turns out that for semisimple groups that will be enough to consider
only constant solutions of (\ref{8}). In the general case the consideration
of the constant solutions of
(\ref{8}) will be inadequate for construction of the generalized Casimir
operator $G$, which is non-degenerate on $M^s$.

A tensor $T$ of type $(p,q)$ on $M^s$ will be the eigenfunction tensor of
the generalized Casimir operator $G$, provided the equation
\begin{equation}\label{13}
        GT \equiv \gamma^{ik} L_{\xi_i} L_{\xi_k} T = \lambda T.
\end{equation}
is satisfied.
This equation can be rewritten in the coordinate form
\begin{equation}\label{14}
{\cal G}T^{a_1...a_p}_{b_1...b_q} \equiv \gamma^{ik}{\cal L}_{\xi_i}{\cal L}_{\xi_k}
T^{a_1...a_p}_{b_1...b_q}=\lambda T^{a_1...a_p}_{b_1...b_q},
\end{equation}
where ${\cal G}T^{a_1...a_p}_{b_1...b_q} \equiv (GT)^{a_1...a_p}_{b_1...b_q}$
are representations of  the generalized Casimir operator acting on the tensor
$T$ in an arbitrary vector basis ${e_a}$ ($e_a(a=1,2,...,s)$ and
$e^a,$~ $e^a(e_b) = \delta^a_b$,  are vector and co-vector bases
on $M^s$ respectively), $T^{a_1...a_p}_{b_1...b_q}$ are the components of the tensor
$T$ with respect to the basis ${e_a}$.

Since the Lie operators "tangle" components of $T$,
it is difficult to solve the equation (\ref{14}) directly. In order to
solve the system of equations (\ref{14}) it is necessary to
"disentangle" components of the tensor $T$ in  the equations, i.e. to
diagonalize the operator $G$. The dioganalization of the generalized
Casimir operator $G$ can be realized invariantly.
Herewith the tensor equations (\ref{14}) split into the system
of scalar differential equations for irreducible components of the tensor
$T$.

Following \cite{Gl-Kon2}, we shall say that
{\bf a split structure} ${\cal H}^s$ is introduced
on $M$ if the $s$ linear symmetric operators (projectors)
$H^a (a=1,2,...s)$ of a constant rank with the properties
\begin{equation}\label{4a}
H^a\cdot H^b=\delta^{ab}H^b; \qquad \sum_{a=1}^{s}H^a=I,
\end{equation}
where $I$ is the unit operator $(I\cdot X=I,~~\forall X\in T(M))$, are
defined on $T(M)$. Here we consider a linear operator $L$
on the tangent bundle $T(M)$ as a tensor of type $(1,1)$ for which
$L\cdot X\equiv L(X)\in T(M),~~\forall X\in T(M)$. The product of two
linear operators $L\cdot H$ obeys the rule $(L\cdot H)\cdot X=L\cdot(H\cdot X)\in T(M),
~\forall X\in T(M)$. An operator $H$ is called a symmetric one if
$(H\cdot X,Y)=(X,H\cdot Y),~\forall X,Y\in T(M)$.

Then we can obtain the decomposition the tangent bundle $T(M)$ and
cotangent bundle $T^{*}(M)$ into the $(n_1+n_2+...+n_s)$ subbundles
$\Sigma^{a}$, $\Sigma_{a}^{*}$, so that
\begin{eqnarray} \label{7a}
    T(M) = {\bigoplus_{a=1}^{s}}\Sigma^{a}; \qquad
T^{*}(M) = {\bigoplus_{a=1}^{s}}\Sigma^{*}_{a}\,.
\end{eqnarray}

Arbitrary vectors, covectors, and metrics are decomposed according
to the scheme:
\begin{equation}\label{8a}
X = \sum^{s}_{a=1} X^a, \quad
\omega = \sum^{s}_{a=1}\omega_a, \quad
g = \sum^{s}_{a=1} g^a, \quad
g^{-1} = \sum^{s}_{a=1} g^{-1}_a
\end{equation}
where
$X^a = H^a\cdot X,~~H^b\cdot X^a = 0,~~X^a\cdot X^b = 0,~~
\omega_a = \omega\cdot H^a,~~\omega_a (X^b) = 0,~~(a\not= b)$.

   Using this scheme we can obtain the decomposition of more complex tensors.

We shall say that a split structure ${\cal H}^s$ is compatible with a
group of isometries if the conditions of invariance of ${\cal H}^s$ are
satisfied, i.e. if
\begin{equation}\label{107}
{L}_{\xi_i }H^a = 0, \quad (i=1,2,...r; a=1,2,...s).
\end{equation}
The equations (\ref{4a}), (\ref{107}) define the invariant projection tensors.

In order to construct the projectors we require existence of such
a dual vector $e_a$ and co-vector $e^b$ bases on $M^s$, that
\begin{equation}\label{17}
e_a\cdot e^b =\delta^b_a~; \quad
H^a=e_a\otimes e^a .
\end{equation}
From now  and on there is no summation on repeating indices $a$ and $b$.
The invariance condition of (\ref{107}) yields
\begin{equation}\label{18}
(L_{\xi_i}e^a)\cdot e_b=0 \quad
(a\! \neq \! b).
\end{equation}
Hence it follows
\begin{equation}\label{19}
        L_{\xi_i}e^a=
\mu^a_ie^a,~~~L_{\xi_i}e_a=-\mu^a_ie_a,
\end{equation}
where the factors of proportionality $\mu^{a}_{i}$ are some functions,
satisfying the equation
\begin{equation}\label{20}
        \xi_i\mu^a_k-\xi_k\mu^a_i=C^j_{ik}\mu^a_j.
\end{equation}
which follows from the integrability condition of (\ref{19}).
Thus, the problem of the construction of the invariant projectors reduces to
construction of the dual vector $\{e_a\}$ and covector
$\{e^b\}$ bases satisfying the corresponding
systems of equations  in (\ref{19}). Some of the factors $\mu^a_i$,
or even all of them in some cases, can vanish. Then the projectors are
constructed by means of the invariant basis $\{e_a: L_{\xi_i}e_a=0\}$.
In particular, for simply transitive groups $(r=s)$,
the invariant vector basis $\{e_a\}$ can be expressed in the form
\begin{equation}\label{22}
        e_a=L^b_a\xi_b \qquad (\det||L^b_a||\neq 0).
\end{equation}
where the factors $L^b_a$ satisfy the equations
$\xi_bL^a_d+C^a_{bq}L^q_d=0$ . The integrability conditions of
these equations are satisfied owing to the Jacobi identity. Using
the last equations it can easily be shown that the tensor $g^{-1}$,
constructed by the formula
\begin{equation}\label{24}
g^{-1}=\delta^{ab}e_a\otimes e_b=g^{ab}\xi_a\otimes \xi_b~;\qquad
g^{ab}=L^a_cL^b_d\delta^{cd}.
\end{equation}
actually satisfies the Killing's equations (\ref{8}), and thereby
will be the inverse metric (\ref{9}).

Now we shall return to the general case of arbitrary bases
$\{e_a,e^b\}$. The initial tensor $T$ can be expanded in the series
\begin{equation}\label{25}
  T=\sum^{ }_{{A,B}}\hat{T}^A_B=\sum^{ }_{{A,B}}T^A_B\hat{e}^B_A,
\end{equation}
where $\{\hat{e}^B_A\}=\{e_{a_1}\otimes\cdots\otimes e_{a_p}\otimes
e^{b_1}\otimes\cdots\otimes e^{b_q}\}$
is the tensor basis, $\hat{T}^{A}_{B} = T^{A}_{B}e^{B}_{A}$
is the tensor monomial and $T^A_B\equiv T^{a_1\ldots a_p}_{b_1\ldots b_q}$
is its component. $A=\{a_1,\ldots,a_p\}$ and $B=\{b_1,\ldots,b_q\}$ are
collective indices. The sum in (\ref{25}) comprises the complete set of indices
${A,B}$. It is easy to show that since the projectors $H_a$ are invariant,
the eigenvalue equations (\ref{13}) and (\ref{14}) split into the set of
independent eigenvalue invariant equations for monomials
\[
G\hat{T}^A_B\equiv \gamma^{ik}L_{\xi_i}L_{\xi_k}\hat{T}^A_B=\lambda\hat{T}^A_B.
\]
Using this relation together with (\ref{19}) and (\ref{20}) we obtain
\begin{equation}\label{26}
{\cal G}T^{A}_{B} \equiv \gamma^{ik}{\cal L}_{\xi_i}{\cal L}_{\xi_k}T^{A}_{B}=
\gamma^{ik}(\xi_i-\phi^{A}_{i B})(\xi_k-\phi^{A}_{k B})T^{A}_{B}=
\lambda T^A_B.
\end{equation}
Here
\begin{equation}\label{27}
\phi^A_{iB} = \sum^p_{k=1}\mu^{a_k}_i - \sum^q_{n=1}\mu^{b_n}_i,~~~
A=\{a_1,\ldots,a_p\},~~~ B=\{b_1,\ldots,b_q\}.
\end{equation}
In order that the tensor equation (\ref{13}) could go over into the
invariantly split equations (\ref{26}) for the irreducible components
$T^A_B$, we must make a change $T\rightarrow T^A_B;~~
L_{\xi_i}\rightarrow{\cal L}_{\xi_i}=\xi_i-\phi^A_{i B}$.
The equations (\ref{26}) can be rewritten in the form
\begin{equation}\label{28}
{\cal G}T^{A}_{B} = [K-2\gamma^{ik}\phi^A_{iB}\xi_k-\gamma^{ik}\xi_i\phi^ A_{kB} +
\gamma^{ik}\phi^A_{iB}\phi^A_{kB}]T^A_B=\lambda T^A_B,
\end{equation}
where $K = \gamma^{ik}\xi_i\xi_k$ is the standard Casimir operator defined in
the space of scalar functions. The solutions of the equations (\ref{28})
and (\ref{19}), (\ref{20}) give us the basic tensor functions
$\hat{T^A_B}=T^A_B\hat{e^B_A}$ in the space of a
tensor representation of the group $G^r$ (or, in other words, tensor
harmonics). Note that if there exists the invariant basis (\ref{22}),
then the generalized Casimir operator (\ref{13}) with respect to
this basis reduces to the standard Casimir operator $K$, and in order to
construct the tensor basis of representation that will be enough to
determine the basis of representation in the space of scalar functions.

We proceed to apply this method to small perturbations in General
Relativity. Let $g^{\mu\nu}$ be the background metric admitting some
group of isometry generated by the set of arbitrary Killing
vectors $\xi_{i}$ ($\cal{L}_{\xi_{i}} g^{\mu\nu} = 0$) , and
$h_{\mu \nu}$ the perturbation. Small perturbations satisfy
the linearized Einstein equations $\delta R_{\mu\nu} = -8 \pi G
\delta S_{\mu\nu}$, where $S_{\mu\nu} = T_{\mu\nu} - (1/2)
g^{\mu\nu} T^\lambda_\lambda$. The perturbation of the Ricci tensor
$\delta R_{\mu\nu}$ satisfies the tensor relation \cite{Winberg}
\begin{equation}\label{W1}
\delta R_{\mu\nu} = \frac{1}{2}g^{\lambda\rho}[h_{\lambda\rho;\mu;\nu}-
h_{\rho\mu;\nu;\lambda}-
h_{\rho\nu;\mu;\lambda}+h_{\mu\nu;\rho;\lambda}].
\end{equation}

The perturbation of the metric and Ricci tensor, being tensors, can be
written with respect to the basis constructed in accordance with
(\ref{18})-(\ref{19}) and extended to the full manifold $M^4$
\begin{equation}\label{W2}
h=h_{ik}e^i\otimes e^k+h_{js}(e^r \otimes e^s
+e^s \otimes e^j)
+h_{ss'}e^s \otimes e^{s'},
\end{equation}
\begin{equation}\label{W3}
\delta R=\delta R^{}_{ik}e^i\otimes e^k+\delta R^{}_{js}(e^r \otimes e^s
+e^s \otimes e^j)
+\delta R^{}_{ss'}e^s \otimes e^{s'}.
\end{equation}
where i,k,j,...are the transversal indices, s,s' denote vectors on the
surface of transitivity on which the underlying group of isometry
acts. The basis $(e^i,e^s) = e^a,$ $(a = i,s = 0,1,2,3)$ we shall call a
$representation~~basis$. The relations
(\ref{W2})-(\ref{W3})
represent the expansion of these tensors in the terms of the irreducible
components $\delta R^{}_{ik}$, $\delta R^{}_{js}$,$\delta R^{}_{ss'}$,
$h_{ik}$, $h_{js}$,
$h_{ss'}$. These components for the perturbation of the
metric can be found by using the above procedure. To find them for the
perturbation of the Ricci tensor we need to write the  tensor relation
(\ref{W1}) with respect to the representation basis. For this purpose one
should substitute the components $h_{\mu\nu}$ for their "irreducible
representatives" $h_{jk}$, $h_{js}$, $h_{ss'}$, and
use the Christoffel symbols $\Gamma_{ab}^{c}$: ($\nabla_{e_a} e_b=
\Gamma_{ab}^{c} e_{c}$) with respect to
the canonical basis when dealing with the covariant
derivatives in (\ref{W1}) (for the formula (\ref{W3}) written out in terms
of the basis vectors and of the Christoffel symbols see Appendix A).
When considering space-times with  matter one should
express the source term $S_{\mu\nu}$ in the irreducible form as
well.

In oder to simplify the form of the perturbations $h_{ab}$ one can
use the gauge freedom, expanding the gauge
vector in terms of the obtained basic functions belonging to the same
eigenvalue as the corresponding perturbations of the metric: $\xi = \xi^{c}
e_{c}$, where $\xi^{c}$, being proportional to the basic
functions, are adjusted to simplify $h_{ab}$. Then the
gauge transformations read
\begin{equation}\label{W3a}
h'_{ab}=h_{ab} + {\cal L}_{\xi^{c} e_{c}} h_{ab}.
\end{equation}

\section{Tensor Harmonics For $G_3$ Groups
         on $M^2$ surfaces of transitivity with
         non-isotropic Killing vectors}

The method, being general, can easily be applied to construction
of tensor harmonics for fields of gravity admitting an arbitrary
group of isometry. We shall consider here one of the most
physically interesting cases, including flat and spherical
symmetries, when a three-parameter group acts on a two dimensional surface
of transitivity.

As is known the orbits $M^2$ of the group of isometry
$G_3$ are  the spaces of
the constant Gauss curvature $\mathbf{K}$, (and the
space-time $M^4$ is of type  either
$D$ or $0$ by Petrov \cite{Petrov2}).
Fields of gravity admitting the $G_3$ group of isometry
acting on the two-dimensional non-isotropic surfaces of transitivity
can be divided into the six canonical types : $G_3 VI$,
$G_3 VII$ ($\mathbf{K} = 0$; the signatures on $M^2$ are $+-$ and $++$
respectively), and
$G_3 VIII$ and $G_3 IX$ (four types) \cite{Petrov1}, \cite{Petrov2}.
From now and
on we shall denote an arbitrary set of
coordinates as $x,y,z$.

\emph{$\mathbf{(a)}$  $G_3 VI$ and $G_3$ $VII$ on $M^2$.}
The Killing vectors have the form:
$\xi_1 =\partial_x$, $\xi_2 =\partial_y$, $\xi_3 = x \partial_y +
\varepsilon y \partial_x$, where $\varepsilon = \pm 1$ for the groups
$G_3 VI$ and $G_3 VII$ respectively.
They can be expressed in a more appropriate form in the cylindrical
coordinates on Euclidean (for $G_3 VII$) and pseudo-Euclidean ($G_3 VI$)
planes which we shall denote in both cases as $r, \varphi$.
\begin{equation}\label{W4}
\xi_1 = -\cos \varphi \partial_r+
r^{-1} \sin \varphi  \partial_{\varphi},\quad
\xi_2 = -\sin \varphi \partial_r-
r^{-1} \cos \varphi \partial_{\varphi},\quad
\xi_3 = -\partial_{\varphi}.
\end{equation}
\begin{equation}\label{W4a}
\xi_1 = -\cosh \varphi\partial_{r}+
r^{-1} \sinh \varphi \partial_{\varphi},\quad
\xi_2 = \sinh \varphi \partial_{r}-
r^{-1} \cosh \varphi \partial_{\varphi},\quad
\xi_3 = -\partial_{\varphi}.
\end{equation}

These groups are integrable and not semi-simple; the corresponding
Cartan tensors are degenerate. The solution of the Killing equations in the
cylindrical coordinates gives
the metric for construction of the Casimir operator: $\gamma^{11} =
-\gamma^{22} = 1$, for $G_3 VI$ and $\gamma^{11} =
\gamma^{22} = 1$, $\gamma^{33} = 0$ for $G_3 VII$.
The basis
\begin{equation}\label{W5}
e_1 = e^{c \varphi} (\partial_r +c r^{-1}
\partial_{\varphi}),
\qquad e_2 = e^{-c \varphi} (\partial_r - c r^{-1}
\partial_{\varphi}),
\end{equation}
where $c=1$ for  $G_3 VI$ and $c=i$ for $G_3 VII$,
can be chosen as the representation one on the surface of transitivity,
for it satisfies (\ref{18})-(\ref{19})
at $\mu_3^1 = -c$, $\mu_3^2 = c$. In this case the construction of
tensor harmonics reduces to that of scalar harmonics. The latter are
the eigenfunctions of the operators:
\begin{equation}\label{W6}
\xi_3 t^{\lambda}_{\mu}=- \partial_{\varphi} t^{\lambda}_{\mu}
=-i \mu t^{\lambda}_{\mu},
\end{equation}
and
\begin{equation}\label{W7}
K t^{\lambda}_{\mu} = (\xi_1^2- \xi_2^2)t^{\lambda}_{\mu}
 = -(\partial^2_{r}+ r^{-1} \partial_{r}-
r^{-2} \partial^2_{\varphi}) t^{\lambda}_{\mu} = -\lambda^2 t^{\lambda}_{\mu}.
\end{equation}
\begin{equation}\label{W7a}
K t^{\lambda}_{\mu} = (\xi_1^2 + \xi_2^2)t^{\lambda}_{\mu}
 = (\partial^2_{r}+ r^{-1} \partial_{r}+
r^{-2} \partial^2_{\varphi}) t^{\lambda}_{\mu} = -\lambda^2 t^{\lambda}_{\mu}.
\end{equation}

From (\ref{W6}),~(\ref{W7}), (\ref{W7a})  it follows $ t^{\lambda}_{\mu} =
(-c)^\mu e^{c\mu \varphi} f^\lambda(r)$
where $f^\lambda(r)$ satisfies the corresponding equations
\begin{equation}\label{W8}
(\partial^2_r + r^{-1} \partial_{r}-\lambda^2-
\mu^2 r^{-2}) f^{\lambda} = 0, \quad
(\partial^2_r + r^{-1} \partial_{r}+\lambda^2-
\mu^2 r^{-2}) f^{\lambda} = 0.
\end{equation}

The solution of the first equation in (\ref{W8}) is the {\it McDonald
function}  $f^{\lambda} =
K_{\mu} (\lambda r)$, and of the second is the {\it cylindrical function}
$f^{\lambda} =  Z_{\mu} (\lambda r)$  \cite{3} .
Thus the scalar harmonics can be expressed in the
form
\[
t^{\lambda}_{\mu} = (-i)^\mu e^{\mu \varphi} \left(C_1 K_{\mu}
(\lambda r) +C_2 K_{-\mu} (\lambda r)\right),
\]
\begin{equation}\label{W9}
t^{\lambda}_{\mu} = (-i)^\mu e^{i\mu \varphi} \left(C_1 Z_{\mu}
(\lambda r) +C_2 Z_{-\mu} (\lambda r)\right).
\end{equation}

It is important for further application that the particular
solutions of the second equation in (\ref{W8})
are the Hankel functions of
the first and second orders. The tensor symmetric harmonic of
weight $\lambda$ and type $(0,2)$ are expressed in the form
\begin{equation}\label{W10}
T^{\lambda \mu}=T^{(\lambda \mu)}_{ik}e^i\otimes e^k+T^{(\lambda \mu)}_{js}(e^j \otimes e^s
+e^s \otimes e^j)
+T^{(\lambda \mu)}_{ss'}e^s \otimes e^{s'},
\end{equation}
where the basic vectors $e^i$ are invariant under the group
transformations, and
the basis $e^s,~(s,s'= 1,2) $ on the surface of transitivity dual to that
defined by (\ref{W5}) is
\begin{equation}\label{W11}
e^1
=  \frac{1}{2} e^{-c\varphi}(d r+  \frac{r}{c} d\varphi),\qquad
e^2
=  \frac{1}{2} e^{c\varphi}(d r-  \frac{r}{c} d\varphi)~.
\end{equation}

The irreducible components can be written in the form
\begin{equation}\label{W12}
T^{(\lambda \mu)}_{ik}=h_{ik}t^{(\lambda \mu)};~~T^{(\lambda \mu)}_{js}=
h_{js} t^{(\lambda \mu)}; \quad
T^{(\lambda \mu)}_{ss'}=h^{(\lambda \mu)}_{ss'}t^{(\lambda \mu)}.
\end{equation}

From the above and the well-known recurrent relations for scalars
\cite{3} we obtain the corresponding relations for tensors.

\emph{$\mathbf{(b)}$  $G_3 VIII$ and $G_3$ $IX$ on $M^2$.}
Here we have the following sets of the Killing vectors:
\begin{eqnarray}
\label{W13}
&&\xi_1 = \cosh y \partial_{x}-\sinh y \coth x
\partial_{y},\quad
\xi_2 = \sinh y \partial_{x}-\cosh y \coth x
\partial_{y},\quad
\xi_3 = \partial_{y}
\\
\label{W14}
&&\xi_1 = -\cos y \partial_{x}+\sin y \coth x
\partial_{y},\quad
\xi_2 = \sin y \partial_{x}+\cos y \coth x
\partial_{y},\quad
\xi_3 = -\partial_{y},
\\
\label{W15}
&&\xi_1 = -\cos y \partial_{ x}+\sin y \cot x
\partial_{y},\quad
\xi_2 = \sin y \partial_{x}+\cos y \cot x
\partial_{y},\quad
\xi_3 = -\partial_{y},
\\
\label{W16}
&&\xi_1 = \cosh y \partial_{x}-\sinh y \cot x
\partial_{y},\quad
\xi_2 = \sinh y \partial_{x}-\cosh y \cot x
\partial_{y},\quad
\xi_3 = -\partial_{y}.
\end{eqnarray}
The groups generated by these operators are semi-simple and non-integrable.
The Cartan tensor is non-degenerate in all these case, and
$\gamma^{ik}= \diag (-1,1,1)$, $\gamma^{ik}= \diag (1,1,-1)$,
$\gamma^{ik}= \diag
(1,1,1)$, $\gamma^{ik}= \diag (-1,1,1)$.

Go over  from $\xi_1$, $\xi_2$ to the creation and annihilation operators
\begin{equation}\label{W17}
H_s=e^{s y}(s\partial_{x}+ \alpha \coth x
\partial_{y}),~~~
\end{equation}
where $s = \pm 1,~\alpha=-1$ and $s = \pm i,~\alpha=-1$
for  the groups generated by the operators
(\ref{W13}) and (\ref{W14}) respectively, and
\begin{equation}\label{W18}
H_s=e^{s y}(s\partial_{x}+ \alpha\cot x
\partial_{y}),
\end{equation}
where $s=\pm \imath,~\alpha=-1$ and
$s = \pm 1,~\alpha=-1$  for  the groups generated by the operators
(\ref{W15}) and (\ref{W16}) respectively.

The operators $L_{H_S}$ are the creation and annihilation operators for tensor
functions $\hat{T^A_B}$, which, in their turn, are the eigenfunctions of
the operator $L_{H_3}$. In the spirit of the book \cite{3}
we can show that there is the set of tensor functions $T^{(l) A}_{(m) B}$
for which
\begin{equation}\label{34}
{\cal L}_{H_3}T^{(l) A}_{(m) B} = mT^{(l) A}_{(m) B}~~(m=-l,-l+1,\ldots,l)~;
\end{equation}
\begin{equation}\label{35}
{\cal L}_{H_S}T^{(l) A}_{(m) B} = \sqrt{l(l+1)-m(m+s)}T^{(l) A}_{(m) B}~,
\end{equation}
where $l$ is the weight of the representation.
Herewith the tensor eigenfunction
equations (\ref{14}) can be written in the form
\begin{equation}\label{36}
-{\cal G}T^{(l) A}_{(m) B}=l(l+1)T^{(l) A}_{(m) B}.
\end{equation}
where $\cal G$ is the generalized Casimir operator.

Suppose that we need to consider the tensor symmetric harmonics of
type $(2,0)$ and of weight $l$, which we shall denote $T^l$. With respect to
the initial differential basis
${e^z,~~dx^1= d x,~~dx^2= d y}$
they can be written in the form
\begin{equation}\label{37}
T^l=T^{(l)}_{zz}e^z\otimes e^z
+T^{(l)}_{ra}(e^z\otimes ~dx^a+dx^a\otimes e^z)
+T^{(l)}_{ab}dx^a\otimes dx^b.
\end{equation}
Note, that the covector $e^z=dz$
is invariant with respect to these groups. In order to split the system of
equations (\ref{36}) for the tensor (\ref{37}), into the irreducible
components, one should go over into the basis of one-forms on the surfaces of
transitivity $M^2$ satisfying the condition (\ref{19}).

It turns out that the covectors
\begin{equation}\label{38}
e^s
=d x + s\sin x d y;~~ e^s=d x + s\sinh x d y;~~(s\in \pm 1,\imath)
\end{equation}
are required for the groups generated by the operators
(\ref{W13}), (\ref{W14}) and (\ref{W15}), (\ref{W16}) respectively.
The condition of invariance of a split structure (\ref{107}) for the
Lie operators associated with the vectors (\ref{W17})
stipulates the  corresponding relations
\begin{equation}\label{39}
\mu^s_{s'}=\frac{ss'e^{is' y}}{\sin x};~~
\mu^s_{s'}=\frac{ss'e^{is' y}}{\sinh x};~~
\mu^s_3=0~~
(s,s'=\pm 1).
\end{equation}
By using (\ref{38}) the relation (\ref{37}) can be rewritten in the form
\begin{equation}\label{40}
T^l=T^{(l)}_{zz}e^z\otimes e^z+T^{(l)}_{zs}(e^r \otimes e^s
+e^s \otimes e^z)
+T^{(l)}_{ss'}e^s \otimes e^{s'}.
\end{equation}
where the sum on $s=\pm 1$ is implied. Then, if we suppose
\begin{equation}\label{41}
T^{(l)}_{zz}=h_{zz}t^l;~~T^{(l)}_{zs}=h_zt^l_s;~~T^{(l)}_{s's}=ht^l_{s+s'},
\end{equation}
where $h_{zz},~h_z,~h$ are functions of $z$, then for the function
$t^l_n~~(n=0,..., s,..., s+s')$ we obtain
\begin{equation}\label{42a}
\left\{\frac{1}{\sinh x}\frac{\partial }{\partial x}\sinh x
\frac{\partial }{\partial x}+\frac{1}{\sinh^2x}
\left(\frac{\partial^2}{\partial y^2}-2 i n \cosh x\frac{\partial}{\partial y}-n^2\right)
+l(l+1)\right\}t^l_n=0.
\end{equation}
for groups generated by the Killing vectors (\ref{W14}),
and exactly the same equation with ordinary $\cos x,$, $\sin x$ instead
of the hyperbolic ones for (\ref{W15}). For the other two cases we
obtain
\begin{equation}\label{42b}
\left\{\frac{1}{\sinh x}\frac{\partial }{\partial x}\sinh x
\frac{\partial }{\partial x}-\frac{1}{\sinh^2x}
\left(\frac{\partial^2}{\partial y^2}-2  n \cosh x\frac{\partial}{\partial y}+n^2\right)
+l(l+1)\right\}t^l_n=0.
\end{equation}
for (\ref{W13}) and exactly the same equation with the corresponding
ordinary trigonometric functions for (\ref{W16}).

Owing to (\ref{34}) we find  the general solutions
for groups generated by the Killing vectors (\ref{W13}),
(\ref{W14}),and (\ref{W15}), (\ref{W16}) respectively
\begin{equation}\label{43}
t^l_n=\Sigma_{m} C_m e^{cm y}{\BB}^l_{nm},
\end{equation}
and
\begin{equation}\label{43a}
t^l_n= \Sigma_{m} C_m e^{cm y}P^l_{nm}
\end{equation}
where $c=1$ for (\ref{W13}), (\ref{W16}) and $c=i$ for
(\ref{W14}), (\ref{W15}), $C_m$ are some coefficients. Here
the function ${\BB}^l_{nm}$ satisfies
the corresponding differential equation following from 
(\ref{42a}):
\begin{equation}\label{44a}
\left\{\frac{1}{\sinh x}\frac{d}{d x}\sinh x
\frac{d}{d x}-\frac{m^2-2mn\cosh x+n^2}{\sinh^2x}
+l(l+1)\right\}{\BB}^l_{nm}=0,
\end{equation}
and $P^l_{nm}$ satisfies exactly the same equation with
ordinary $\cos x,$, $\sin x$.
The solutions of the obtained equations are the functions
${\BB}^l_{nm}(\cosh x)$ and $P^l_{nm}(\cos x)$ and ,
which are called the {\it Legendre
functions} and {\it Legendre polynomials}
respectively \cite{3}.
Recurrent relations for them follow from (\ref{34}),(\ref{35}).

It is obvious that in a similar fashion we can treat all the other
types of gravity in Petrov classification scheme.

\section{Applications}
\subsection{Weak cylindrical gravitational waves as perturbations
         of the flat space-time}

Now as a quick illustration of the previous section we shall consider
the background metric of a flat space-time in the form
\begin{equation}\label{W30}
ds^2= d t^2-d r^2-r^2 d \varphi^2-d z^2.
\end{equation}
Except $\partial_t$  $\partial_z$ it has the Killing's vectors (\ref{W4})
and thus can be expanded with the help of  the cylindrical
harmonics (\ref{W9}). The representation  basis of the full space-time
according to (\ref{W5}) is
\begin{equation}\label{W33a}
e_0 = \partial_t,~~e_1 = e^{i \varphi}(\partial_r +
\frac{i}{r}\partial_\varphi),~~e_2 = e^{-i \varphi}(\partial_r -
\frac{i}{r}\partial_\varphi),~~e_3 = \partial_z
\end{equation}
Then according to (\ref{A1}) we almost immediately
obtain  the irreducible combinations of the linearized Einstein
equations (\ref{W34} - \ref{W40})
writing the perturbations of the Ricci tensor with respect to the
representation basis. Since these equations are
invariant under the change of a sign of time, i.e. since the
parity operator commutes with all the other operators of the
underlying group one can write the perturbations $h_{\mu \nu}$
as the sum of parts associated with positive and negative frequencies
$\pm \omega = \pm \sqrt{k^2+\lambda^2}$:
\begin{equation}\label{W31}
h_{ab}(x)= h_{ab}^{+}(x)+h_{ab}^{-}(x),
\end{equation}
where $h_{ab}^{\pm}(x)$ are expanded into the series
\begin{equation}\label{W32}
h_{ab}^{\pm}(x)= \int_{-\infty}^{+\infty}\exp (\pm i \omega t
\mp i k z) d k \sum_{n = -\infty}^{n = +\infty}\int_{0}^{+\infty}
a_{ab}^{\pm}(x)\Phi_{n}^{\pm} \lambda d \lambda.
\end{equation}
Here $\Phi_{n}^{\pm}$ being an eigenfunction of the Casimir
operator with the eigenvalue $-\lambda^2$ satisfies the relations
\begin{equation}\label{W33}
K \Phi_{n}^{\pm} = -\lambda^2 \Phi_{n}^{\pm},~~~
e_{1}\Phi_{n}^{\pm} = \pm i \lambda \Phi_{n \pm 1}^{\pm},~~~
e_{2}\Phi_{n}^{\pm} = \pm i \lambda \Phi_{n \mp 1}^{\pm},~~~
\Phi_{n}^{+}=(\Phi_{n}^{-})^{\ast}
\end{equation}
where $e_{1}$, $e_{2}$ are determined by (\ref{W5}), and $K$ by
(\ref{W7a}).

Then  by analogy with electrodynamics if we choose the Bessel function
as the solution of (\ref{W8}) we easily find
the system of algebraic equations (see Appendix (\ref{c1} - \ref{c4}))
describing the standing waves in
the $r-\varphi$ plane \cite{Radchen}, and if the Hankel function -
the running cylindrical waves. These equations are equivalent to
the coordinate conditions
\begin{equation}\label{W33b}
\Gamma_{\alpha \beta \gamma} g^{\alpha \beta} = \frac{1}{r}
h_{1 \gamma},
\end{equation}
where all the objects are determined with respect to the
cylindrical coordinates basis $d t, d r, d \varphi, d z$ (remember
that $\alpha, \beta, \mu, \nu,...$ are the coordinates indices).
 Then we can use the gauge freedom
to decrease the number of independent coefficients $(a_{ab})_n$.

\subsection{Spherical symmetry: Regge - Wheller harmonics}

   Consider the background metric in the Schwarzshild form
\begin{equation}\label{W41}
d s^2 = - (1-\frac{2m^\ast}{r})d t^2 + (1-\frac{2m^\ast}{r})^{-1}d r^2
+ r^2(d \theta^2 + sin^2\theta d \varphi^2)
\end{equation}
The Schwarzshild space-time being  stationary and spherically symmetric
admits the four Killing vectors:  the three defined in (\ref{W15})
and $\partial_t$. Hence it follows that the basis
\begin{equation}\label{W42}
e_0 = \partial_t,~~~
e_1 = \partial_r,~~~
e_2 = \frac{1}{2}(\partial_\theta -
 \frac{i}{\sin \theta} \partial_\varphi),~~~
e_3 = \frac{1}{2}(\partial_\theta +
 \frac{i}{\sin \theta} \partial_\varphi).
\end{equation}
can be chosen as a representation basis.
The Christoffel symbols evaluated at the pole with respect to this
basis are given by the formulas

Substituting  The Christoffel symbols evaluated at the
pole with respect to this basis   for those in (\ref{A1})
and using expansion into spherical
harmonics $P_{mn}^l$ at $m=0$ after the  Regge-Wheeler  gauge
transformations in the representation basis (\ref{W42})
we can write the perturbations of the Ricci tensor with respect
to this basis, i.e.in the form of the irreducible combinations of
the linearized Einstein equations $\delta R_{a b} = 0$,
and thus separate radial variable and time from the angular
variables.
The eventual equations are equivalent to
those of the works  \cite{Reg} and \cite{Zeril2} and
therefore are not listed here. The basis (\ref{W42}), that can be
multiplied by an arbitrary function of $r$ and $t$, are widely
used (see for example  \cite{Kalnins}) and the tensor
harmonics following from it were discussed by many authors \cite{Zeril1},\cite{Hu} .

\section*{Acknowledgement}
I would like to acknowledge Professor V.D.Gladush for proposing
the problem, much encouragement and reading the manuscript.
I am also much indebted to Dr. S.Ulanov for computer help.

\appendix
\section*{Appendix}
\section{The linearized Einstein equations for empty space
with respect to arbitrary basis}

The relation (\ref{W3}) written out in terms
of the Christoffel symbols and of the basis vectors reads
\begin{eqnarray}
2\delta R_{ab} &=& g^{ab} \left[ e_a e_b h_{mn} - e_n e_b h_{ma} -
e_m e_b h_{na} + e_b (( \Gamma^{l}_{mn} +
\Gamma^{l}_{nm})h_{la} \right. \nonumber \\
&&+ ( \Gamma^{l}_{an} - \Gamma^{l}_{na})h_{lm} +
( \Gamma^{l}_{am} - \Gamma^{l}_{ma})h_{ln}) + \Gamma^{l}_{ab}
(e_n h_{ml} +  e_m h_{nl} - e_l h_{mn}) \nonumber \\
&&+\Gamma^{l}_{mb}
(e_n h_{la} +  e_l h_{na} - e_a h_{ln}) -
(\Gamma^{k}_{mb}(\Gamma^{l}_{kn} + \Gamma^{l}_{nk}) +
\Gamma^{k}_{nb}(\Gamma^{l}_{mk} + \Gamma^{l}_{km}))h_{la}+ \nonumber \\
&& + \Gamma^{k}_{nb}
(e_m h_{ka} + e_k h_{ma} - e_a h_{km}) +
(\Gamma^{k}_{mb}(\Gamma^{l}_{ka}-\Gamma^{l}_{ak}) + \Gamma^{k}_{ab}
(\Gamma^{l}_{km} - \Gamma^{l}_{mk}))h_{ln} \nonumber \\
&&\left. -\Gamma^{k}_{ab}(\Gamma^{l}_{mn} + \Gamma^{l}_{nm}) + (\Gamma^{k}_{ab}
(\Gamma^{l}_{nk}-\Gamma^{l}_{kn}) + \Gamma^{k}_{nb}(\Gamma^{l}_{ka}-
\Gamma^{l}_{ak}))h_{ml}\right] + e_m e_n h - \Gamma^{l}_{mn} e_l h=0,\label{A1}
\end{eqnarray}
where $a,b,m,n,k,l,...=0,1,2,3$ are indices associated with the
basis vectors $e_0,e_1,...$, so that $g_{ab},~\Gamma^{a}_{bc},$ and $h_{ab}
=h_{ba}$
are the background metric, the Christoffel symbols, and the perturbations
determined with respect to the basis $e_a~(a=i,s=0,1,2,3)$.

\section{The linearized Einstein equations for
cylindrical waves as perturbations of the flat space-time in the
form of the irreducible combinations}
\begin{eqnarray}
\label{W34}
\delta R_{00} &=& -\frac{1}{2}(e_1 e_2 + e_3 e_3)h_{00} + \frac{1}{2}
e_0 e_2 h_{01} + e_0 e_1 h_{02} + e_0 e_3 h_{03} - e_0 e_0 (h_{12} +
h_{33}), \\
\delta R_{01} &=& -\frac{1}{2}(-\frac{1}{2}e_1 e_2 - e_3 e_3)h_{01}
+ \frac{1}{4} e_1 e_1 h_{02}
+ \frac{1}{2} e_1 e_3 h_{03} +
\frac{1}{4} e_0 e_2 h_{11}
- \frac{1}{4}e_0 e_1 h_{12}
\nonumber \\
&&+\frac{1}{2} e_0 e_3
h_{13} - \frac{1}{2}e_0 e_1 h_{33} = 0, \label{W35}
\\
\label{W36}
\delta R_{03} &=& \frac{1}{4}e_3 e_2 h_{01} + \frac{1}{4}e_3 e_1
h_{02}- \frac{1}{2}e_1 e_2 h_{03} - \frac{1}{2}e_0 e_3 h_{12} +
\frac{1}{4}e_0 e_2 h_{13} + \frac{1}{4}e_0 e_1 h_{23} = 0, \\
\delta R_{13} &=& \frac{1}{2}e_1 e_3 h_{00} - \frac{1}{2}e_0 e_3
h_{01}- \frac{1}{2}e_0 e_1 h_{03} +
\frac{1}{4}e_3 e_2 h_{11} -
\frac{1}{4}e_1 e_3 h_{12} + \frac{1}{2}(e_0 e_0 - \frac{1}{2}e_1
e_2)h_{13} \nonumber \\
\label{W37}
&&+ \frac{1}{4}e_1 e_1 h_{23} = 0, \\
\delta R_{12} &=& \frac{1}{2}e_1 e_2 h_{00} - \frac{1}{2}e_0 e_2
h_{01}- \frac{1}{2}e_0 e_1 h_{02} + \frac{1}{4}e_2 e_2 h_{11} +
\frac{1}{2}( e_0 e_0 - e_1 e_2 - e_3 e_3)h_{12} + \frac{1}{2}
e_2 e_3 h_{13} \nonumber \\
\label{W38}
&&
+ \frac{1}{4}e_1 e_1 h_{22} + \frac{1}{2}e_1 e_3 h_{23}
-\frac{1}{2}e_1 e_2 h_{33} = 0, \\
\label{W39}
\delta R_{11} &=& \frac{1}{2}e_1 e_1 h_{00} - e_1 e_0
h_{01}+ \frac{1}{2}(e_0 e_0 -e_3 e_3)h_{11}+e_1 e_3 h_{13}-
\frac{1}{2}e_1 e_1 h_{33} = 0, \\
\delta R_{33} &=& \frac{1}{2}e_3 e_3 h_{00} - e_0 e_3
h_{03}- \frac{1}{2}e_3 e_3 h_{12} +\frac{1}{2}e_3 e_2 h_{13} +
\frac{1}{2} e_3 e_1 h_{23} + \frac{1}{2}
(e_0 e_0 - e_1 e_2)h_{33} = 0.\label{W40}
\end{eqnarray}

\section{The algebraic equations describing the cylindrical
waves}
\begin{equation}\label{c1}
\omega(a_{00})_{n} + \lambda[(a_{01})_{n+1} + (a_{02})_{n-1}] +
2 k (a_{03})_{n} + \omega (a_{12})_{n} + \omega (a_{33})_{n} = 0,
\end{equation}
\begin{equation}\label{c2}
\lambda [(a_{33})_{n} - (a_{00})_{n} -(a_{11})_{n+2}] - 2 \omega(a_{01})_{n+1}
- 2 k (a_{13})_{n+1} = 0,
\end{equation}
\begin{equation}\label{c3}
\lambda [(a_{33})_{n} - (a_{00})_{n} -(a_{22})_{n-2}] - 2 \omega(a_{02})_{n-1}
- 2 k (a_{23})_{n-1} = 0,
\end{equation}
\begin{equation}\label{c4}
\lambda [(a_{00})_{n} - (a_{12})_{n} +(a_{33})_{n}] + 2 \omega(a_{03})_{n}
+ \lambda[(a_{13})_{n+1} + (a_{23})_{n-1}] = 0,
\end{equation}
 $$a^{+}_{ab}(k,n,\lambda) \equiv (a_{ab})_n.$$

\def\CMPh{Commun. Math. Phys.}
\def\JPh{J. Phys.}
\def\CJP{Czech. J. Phys.}
\def\LMPh {Lett. Math. Phys.}
\def\NPh  {Nucl. Phys.}
\def\PhE  {Phys.Essays}
\def\PhL  {{\it Phys. Lett.}~}
\def\PhR  {{\it Phys. Rev.}~}
\def\PhRL {Phys. Rev. Lett.}
\def\PhRp {Phys. Rep.}
\def\NCim {Nuovo Cimento}
\def\NuPB {Nucl. Phys.}
\def\GRG {{\it Gen. Relativ. Gravit.}~}
\def\CQG {Class. Quantum Grav.}
\def\prp {report}
\def\Prp {Report}
\def\GrC {{\it Gravitation$\&$Cosmology}~}
\def\DANS {{\it Dokl.Akad.Nauk SSSR}~}
\def\APh {{\it Ann.Phys.}~}
\def\JMM {{\it Journ.Math. and Mech.}~}
\def\JMP {{\it J.Math. Phys.}~}
\def\IVUZ {{\it Izv.Vyssh.Uchebn.Zaved.Fiz}~}
\def\APP {{\it Acta Phys.Pol.}~}

\end{document}